\documentclass[prl,aps,twocolumn,amsmath,amssymb]{revtex4}
\usepackage{graphicx}
\usepackage{dcolumn}

\begin{document}

\title{Quantum Phase Transition in a Resonant Level Coupled to Interacting Leads}

\author
{Henok  T. Mebrahtu,$^{1}$ Ivan V.  Borzenets,$^{1}$ Dong E. Liu,$^{1}$ Huaixiu Zheng,$^{1}$ \\
Yuriy V. Bomze,$^{1}$ Alex I. Smirnov,$^{2}$ Harold U. Baranger,$^{1}$ and Gleb Finkelstein$^{1}$}
\affiliation{$^{1}$ Department of Physics, Duke University, Durham, NC 27708}
\affiliation{$^{2}$ Department of Chemistry, North Carolina State University, Raleigh, NC 27695}

\begin{abstract}
An interacting one-dimensional electron system, the Luttinger liquid, is distinct from the ``conventional'' Fermi liquids formed by interacting electrons in two and three dimensions \cite{Giamarchi_1}. Some of its most spectacular properties are revealed in the process of electron tunneling: as a function of the applied bias or temperature the tunneling current demonstrates a non-trivial power-law suppression \cite{Albert_1, Deshpande_1}. Here, we create a system which emulates tunneling in a Luttinger liquid, by controlling the interaction of the tunneling electron with its environment. We further replace a single tunneling barrier with a double-barrier resonant level structure and investigate {\it resonant} tunneling between Luttinger liquids. For the first time, we observe perfect transparency of the resonant level embedded in the interacting environment, while the width of the resonance tends to zero. We argue that this unique behavior results from many-body physics of interacting electrons and signals the presence of a quantum phase transition (QPT) \cite{Sachdev_1}. In our samples many parameters, including the interaction strength, can be precisely controlled; thus, we have created an attractive model system for studying quantum critical phenomena in general. Our work therefore has broadly reaching implications for understanding QPTs in more complex systems, such as cold atoms \cite{Bloch_1} and strongly correlated bulk materials \cite{Si_1}.
\end{abstract}


\maketitle

Unlike two- and three-dimensional Fermi liquids, a Luttinger liquid completely ``dissolves'' individual electrons, replacing them with collective plasmon waves. When in the process of quantum-mechanical tunneling an outside electron is added to the Luttinger liquid, the plasmons spread the charge through the system, akin to the ripples from a raindrop on the surface of a pond. At zero temperature, the tunneling electron does not have the necessary energy to excite the plasmons. As a result, the tunneling conductance between a normal metal and a Luttinger liquid, or between two Luttinger liquids, is suppressed at low temperature as a power law \cite{Albert_1,Deshpande_1}. 

Even more interesting is the case of resonant tunneling, in which a single tunnel barrier between Luttinger liquids is replaced by a \emph{resonant} level formed in a double-barrier quantum structure. Starting with the seminal papers by Kane and Fisher \cite{KF_1}, this problem has received significant theoretical attention \cite{Eggert&Affleck1992_1,NG_1,PG_1,KG_1}. Perhaps the most spectacular prediction is the existence of resonance peaks with perfect conductance (full transparency), but vanishingly small width (infinite life time) at zero temperature \cite{KF_1}. These resonances require two identical Luttinger liquids that are symmetrically coupled to the resonant level. Although several experiments addressed resonant tunneling in a Luttinger liquid \cite{Milliken_1,Maasilta_1,RT_LL_1}, controlling the tunneling strength was never attempted. 

\begin{figure}[h]
\includegraphics[width=1 \columnwidth]{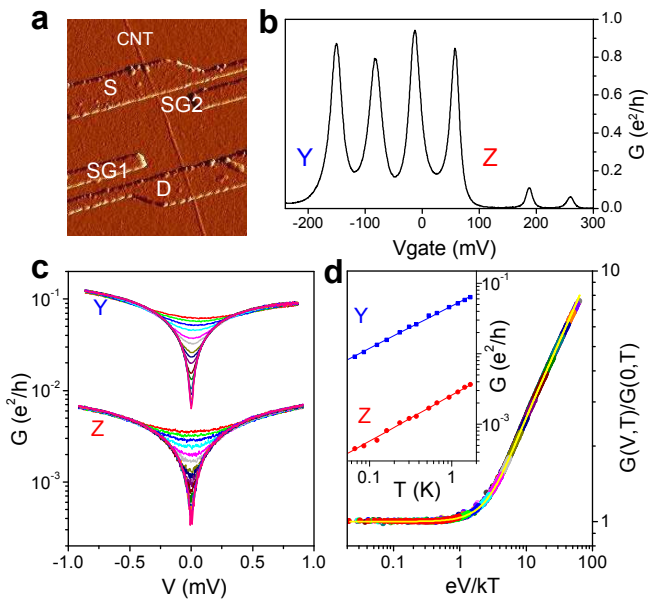}
\caption{\label{fig:overview}
{\bf Emulating Luttinger liquid with resistive environment.} {\bf a,} AFM image of the sample. The nanotube (marked CNT) is contacted by two metal leads (S and D), forming a quantum dot. 
Two side-gates (SG$_1$ and SG$_2$) control the coupling of the dot to the two leads, as used later in Fig.~2. {\bf b,} Differential conductance $G \equiv dI/dV$ of a similar sample as a function of the back-gate voltage $V_{\rm gate}$. 
We focus on Coulomb blockade valleys Y and Z, in which electron transport is conducted through co-tunneling processes. $T= 1.8$ K. {\bf c,} Differential conductance measured \emph{ vs.} bias $V$ demonstrates a pronounced zero-bias anomaly in both valleys. Temperature: 1.7 K to 30 mK (top to bottom). {\bf d,} Inset: zero-bias conductance $G(0, T)$ clearly shows a power law dependence on $T$ with the same exponents found in both valleys. Main: $G(V, T)$ data measured in valley A at different temperatures can be rescaled to collapse on the same universal curve, described by the theoretical expression of Ref.  \cite{Averin_1} (yellow line). 
}
\end{figure}

In this work, we create a system analogous to a Luttinger liquid by properly designing electron interactions in the resonant level's environment and, for the first time, tune the system to the symmetric coupling point. In marked contrast with suppressed low-temperature conductance through a single tunnel barrier, we find that a resonant level, symmetrically coupled to two leads, retains a \emph{unitary} conductance of $e^2/h$, corresponding to perfect transparency. Simultaneously, the width of the resonance tends to zero as a power law of temperature. We argue that this unique behavior results from many-body physics of interacting electrons and signals the presence of a quantum phase transition (QPT) \cite{Sachdev_1,Vojta_1}. QPTs found in strongly correlated bulk materials are often explained by invoking interactions between local sites and collective modes \cite{Sachdev_1}. In the same spirit, our observation provides the first example of a QPT in a highly tunable system, which emulates such a ``local site'' (\emph{i.e.} the resonant level) embedded in an \emph{interacting host}. 

Both the single barrier tunneling and the double-barrier resonant tunneling regimes are realized in short ($\sim 300$ nm) segments of carbon nanotubes (Fig.~1a). The representative electrical conductance through such a sample is shown in Fig.~1b. The size quantization of electron states in the nanotube, combined with the mutual repulsion of the electrons, result in a ``Coulomb blockade'' pattern \cite{Kastner_1,QDreview_1}.

We first show Luttinger liquid-like properties in tunneling through a single barrier by tuning the gate voltage to Coulomb blockade valleys Y and Z of Fig.~1b, where no low-energy excitations exist in the nanotube. Electrons are then transmitted through by the co-tunneling processes (see discussion in the Supplementary Information). These processes are almost energy-independent on energy scales smaller than the nanotube charging energy and level spacing (both meV's) -- the nanotube should behave just like a single tunnel barrier. 

The conductance in valleys Y and Z measured \emph{ vs.} bias $V$ shows a surprising zero-bias anomaly (ZBA), which gets progressively deeper as the temperature decreases (Fig.~1c). Since the shape of the ZBA in the two valleys is the same up to an overall scale factor, the existence of the ZBA is not due to the nanotube itself. Indeed, the distinct feature of our samples is the metal leads to the nanotube, which are made rather resistive (k$\Omega$'s). ``Tunneling with dissipation'' \cite{LeggetRevModPhys_1} between such resistive leads is known to result in suppressed conductance $dI/dV \equiv G \propto max(k_B T, eV)^{2r}$  \cite{NI_1,Flensberg_1,Joyez_1,Averin_1}. This expression is similar to the power-law suppression expected for tunneling in a Luttinger liquid; however no real Luttinger liquid is present in our sample \cite{note1_1}, and the exponent $r=e^2 R /h$ is determined by the ratio of the resistance of the leads $R$ to the quantum resistance $h/e^2$. Note the highly unusual appearance of the resistance in the exponent, which allows us to control the strength of tunneling suppression simply by changing $R$. 

Experimentally, the zero-bias conductance scales as $G(0,T) \propto T^{2r}$, with the same exponent $2r \sim 0.6$ found in both valleys (inset to Fig.~1d); this value is consistent with the leads resistance ($R \approx 6.5$ k$\Omega$ in this sample). Furthermore, we can rescale the whole set of $G(V,T)$ curves measured in valley Y as shown in Fig. 1d, which presents $G(V,T) / G(0,T)$ as a function of $eV/k_B T$ -- a dimensionless ratio of bias to temperature. The yellow curve overlaying the symbols is the result of the full theoretical expression describing tunneling with dissipation \cite{Averin_1}, in which we use the same value of $r = 0.3$ extracted from the temperature dependence. 

The expression used to fit the data in Fig.~1d is identical to the one describing tunneling between two Luttinger liquids \cite{Sassetti_Napoli_Weiss_95_1}. The similarity may be understood qualitatively: both for tunneling in a dissipative environment and in a Luttinger liquid, the tunneling electron's charge couples to a continuum of bosonic modes (plasmons); at zero energy (temperature or bias) the electron cannot excite the modes, and tunneling is suppressed. Furthermore, the formal mapping of the two problems has been demonstrated for the single-barrier case in Ref.  \cite{Safi&Saleur_1}. The recipe is to replace the Luttinger interaction parameter $g$ by $1/(r+1)$: \emph{e.g.} the case of vanishing dissipation, $r = 0$, corresponds to the non-interacting Luttinger liquid, $g = 1$. It is important to realize that for $r \neq 0$ electrons do in fact interact with each other through their coupling to the bosonic modes. We use the analogy between tunneling in a dissipative environment and tunneling in a Luttinger liquid through the rest of this text \cite{note1_1}.

Having established Luttinger liquid-like behavior in a single barrier tunneling, we now turn to the main focus of this paper: resonant tunneling between interacting leads. We study single-electron conductance peaks, similar to those shown in Fig.~1b, but measured on a different sample. A key feature of our experiment is the use of additional side gates to tune the coupling of the resonant level to the leads (Fig.~1a). Fig.~2a shows the differential conductance map as a function of the side and back gate voltages. Clearly, the heights of the peaks change along the traces. For several of the peaks, the conductance reaches a maximum at some intermediate value of the side gate voltage, indicating that the tunneling rates from the resonant level to the source and the drain are equal: $\Gamma_S = \Gamma_D$ (``symmetric coupling'').

We focus on peak {\bf X} of Fig.~2a, with the side gate tuned so that the tunneling is either symmetric (Fig.~2b), or rather asymmetric (Fig.~2c). Clearly, the two cases behave in markedly different ways: in the asymmetric case, the peak height decreases at low temperatures, while the width saturates \cite{dissipation_1}. In the symmetric case, the peak width decreases, while the peak height grows and reaches $e^2/h$  \cite{note0_1}. \emph{It is remarkable that the resonant tunneling conductance can reach the unitary limit despite coupling to the interacting leads, which suppress tunneling in the single barrier case.}

\begin{figure}
\centering
\includegraphics[width=0.7 \columnwidth]{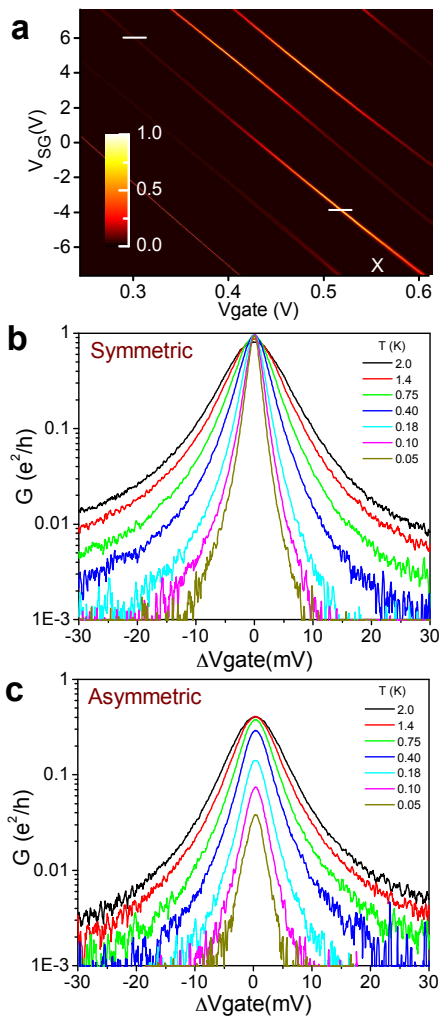}
\caption{\label{fig:main}
{\bf Resonant lineshape: symmetric and asymmetric cases.} {\bf a}, Zero-bias differential conductance as a function of $V_{\rm gate}$ and the voltage $V_{\rm SG}$ applied to one of the side gates. Several peaks reach a maximal conductance of $e^2/h$ along their traces in the range shown here. The base temperature is $T= 50$ mK; perpendicular magnetic field of 6 T is applied to select a single spin species. {\bf b, c,} Resonant conductance measured on the peak marked `X' in panel a at several temperatures as a function of $\Delta V_{\rm gate}$, gate voltage relative to the center of the peak. ({\bf b}) symmetric and ({\bf c}) asymmetric coupling cases (the side gate voltages are fixed at the values indicated by white lines in Fig.~2a). As the temperature is reduced, in the symmetric case the peak becomes taller and narrower; in contrast, in the asymmetric case the peak becomes shorter and its width saturates. 
}
\end{figure}

To account for the observed behavior, in the supplementary material we present a model of a resonant level connected to two electron reservoirs, with excitation of environmental modes represented by a dynamical phase associated with the tunneling matrix element \cite{NI_1}. We show that the analogy between tunneling with dissipation and tunneling in a Luttinger liquid \cite{Sassetti_Napoli_Weiss_95_1,Safi&Saleur_1,LeHur&Li_05_1,Florens_07_1} further extends to our case of {\it resonant tunneling}. Based on our mapping and the Luttinger liquid predictions \cite{KF_1,Eggert&Affleck1992_1,NG_1,PG_1,KG_1}, in the case of symmetric coupling we expect the peak height to saturate at $e^2/h$ (spinless case), and the resonance width to scale at low temperatures $\propto T^{r/(r+1)}$  \cite{KF_1}. For asymmetric coupling, the resonance width is predicted to saturate at low enough temperature \cite{NG_1,PG_1}, while the peak height should scale to zero as $G\propto T^{2r}$, featuring the same exponent as in the single barrier (non-resonant) tunneling.

Our experiment clearly corroborates these predictions (Fig.~3). Quantitatively, we extract $r \approx 0.75$ from the scaling of the asymmetric peak height, which agrees with the leads resistance in this sample. The width of the symmetric peak scales with an exponent of 0.45, consistent with $r/(r+1) \approx 0.43$. (We discuss the accuracy of extracting the exponents in the supplementary.) Overall, application of Refs. \cite{KG_1,Eggert&Affleck1992_1,NG_1,PG_1} to our experiment describes the observed behavior remarkably well, both in the symmetric and asymmetric cases. 

\begin{figure}
\includegraphics[width=0.9 \columnwidth]{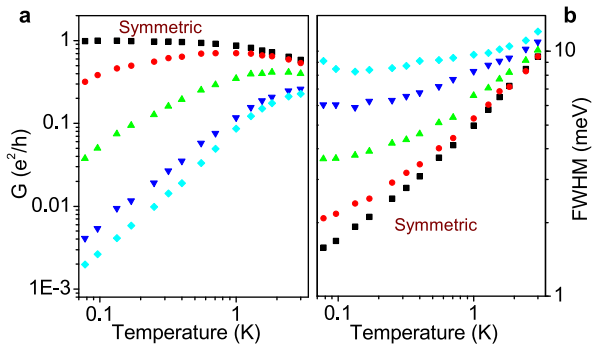}
\caption{\label{fig:H&W}
{\bf Resonant peak parameters at different degrees of asymmetry.} Conductance peak height ({\bf a}) and width ({\bf b}) measured at several values of $V_{\rm SG}$, which controls the degree of the tunnel barrier asymmetry. (Different peak, but the same sample as in Figure 2.) Note that in the symmetric case, the peak height saturates at $e^2/h$, while the peak width monotonically decreases with lowering temperature. In the asymmetric case, the behavior is the opposite: the width of the peak saturates, while the peak height decreases. 
}
\end{figure}

Note that the width of the conductance peak in the symmetric case monotonically decreases with lowering temperature. In the limit of zero temperature, we expect that the conductance will be equal to zero everywhere, except for a singular point at the center of the peak. When tunneling asymmetry is introduced, the singular point disappears, and the low-temperature conductance tends to zero at any $V_{\rm gate}$. This behavior indicates a quantum phase transition \cite{Vojta_1} for symmetric coupling, $\Gamma_S = \Gamma_D$. 

Indeed, in the supplementary material, we map our model in the $r=1$ case, following Ref. \cite{KG_1}, onto the exotic two-channel Kondo model \cite{Hewson_1,DGG_2CK_1}, for which a QPT is known to occur exactly for symmetric coupling \cite{Eggert&Affleck1992_1}. In both models, the origin of the quantum critical behavior is the competition between the two channels attempting to screen the local site (spin or resonant level).

\begin{figure}
\includegraphics[width=1 \columnwidth]{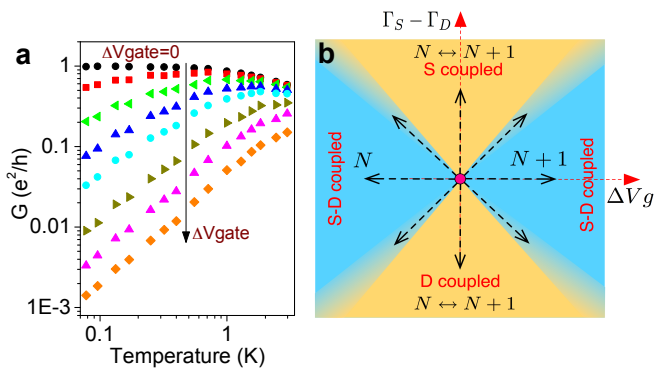}
\caption{\label{fig:VgT}
{\bf Phase diagram and the quantum critical point.} {\bf a,} Conductance in the symmetric coupling case, plotted as a function of temperature at different values of gate voltage, starting with $\Delta V_{\rm gate}=0$ (top curve). Note the similarity to Figure 3a. {\bf b,} Proposed phase diagram: the quantum critical point at the center (symmetric coupling and $\Delta V_{\rm gate}=0$) has unitary conductance. Any deviations from this point result in vanishing conductance at $T=0$.
}
\end{figure}

The intermediate values of $1>r>0$ do not allow for a simple interpretation in terms of any Kondo model with non-interacting leads, but represent a continuous evolution between the non-interacting resonant level at $r=0$ and the two-channel Kondo model \cite{Affleck_1,goldstein10_1}. The critical exponents describing the system parameters close to the quantum critical point are not fixed, but are controlled by the value of $r$. Thus our system not only provides new insights into the two-channel Kondo model -- a paradigmatic example of quantum criticality -- but also gives access to a new family of quantum critical points for $r>0$.

The QPT observed here is different from the various QPTs observed \cite{C60QPT_1} and predicted \cite{LeHur&Li_05_1,Borda_06_1} in quantum dots coupled to a \emph{single} screening channel -- indeed, there the QPTs are of the Kosterlitz-Thouless-type, while in our case the QPT is of second order (see the Supplementary Information). Furthermore, in our case, the key ingredient that enables the QPT is the symmetric coupling to the two leads, which allows for their competition; the interaction in the leads (finite $r$) prevents their hybridization. 

Finally, the conductance in the symmetric case can be plotted as a function of temperature for several values of $\Delta V_{\rm gate}$ (Fig. 4a). The similarity with Fig. 3a is striking: apparently, one can tune away from the unitary resonance either by inducing asymmetry (Fig. 3a) or by applying the gate voltage (Fig. 4a)  \cite{KF_1} with virtually the same results. Note that the downturn of peak height in Figures 3a and 4a occurs at increasingly lower temperature as either the degree of asymmetry or $\Delta V_{\rm gate}$ is reduced. Clearly, a new energy scale is emerging in the system, controlled by proximity to the quantum-critical point  \cite{KF_1,NG_1,PG_1}. We anticipate that this scale should vanish exactly at that point. We therefore propose the phase diagram of Fig. 4b, with a quantum critical point at $\Gamma_S = \Gamma_D$, $\Delta V_{\rm gate}=0$   \cite{Affleck_1}. The four quadrants represent the states of the nanotube filled with $N$ or $N+1$ electrons, or coupled more strongly to either the source or the drain. The boundaries between the quadrants are smeared, and at $T=0$ the conductance tends to zero everywhere, except for the quantum critical point.

In conclusion, we have investigated resonant tunneling between interacting leads emulating Luttinger liquids. For symmetric coupling of the spinless resonant level to the two leads, and on-resonance, the low-temperature conductance saturates at the unitary value of $e^2/h$. We associate this behavior with a quantum critical point, which exists at $\Gamma_S = \Gamma_D$ in the presence of a finite interaction strength $r>0$. Moving away from this point by inducing tunneling asymmetry results in suppression of conductance at low temperature and smearing of the QPT. Our work is the first example of a QPT in a highly tunable system, in which many parameters can be controlled, including the strength of interactions.

\bigskip
We appreciate valuable discussions with I. Affleck, D.V. Averin, A.M. Chang, C.H. Chung, S. Florens, M. Goldstein, L.I. Glazman, K. Ingersent, K. Le Hur, M. Lavagna, A.H. MacDonald, Yu.V. Nazarov, D.G. Polyakov, and M. Vojta. We thank J. Liu for providing the nanotube growth facilities and W. Zhou for helping to optimize the nanotube synthesis. The work was supported by U.S.\,DOE awards DE-SC0002765, DE-SC0005237, and DE-FG02-02ER15354.

\newpage
\begin{widetext}

\begin{center}
{\large {\bf Supplementary Information for \\
``Quantum Phase Transition in a Resonant Level Coupled to Interacting Leads''}\\
\bigskip
Henok  T. Mebrahtu,$^{1}$ Ivan V.  Borzenets,$^{1}$ Dong E. Liu,$^{1}$ Huaixiu Zheng,$^{1}$ \\
Yuriy V. Bomze,$^{1}$ Alex I. Smirnov,$^{2}$ Harold U. Baranger,$^{1}$ and Gleb Finkelstein$^{1}$}\\
\medskip
$^{1}$ Department of Physics, Duke University, Durham, NC 27708\\
$^{2}$ Department of Chemistry, North Carolina State University, Raleigh, NC 27695

\end{center}
\end{widetext}

\setcounter{figure}{0}
\renewcommand{\thefigure}{S\arabic{figure}}

\section{Nanotube quantum dots}

The nanotubes are grown by chemical vapor deposition from a CH$_4$ feedstock gas \cite{HongjieDai} on a Si/SiO$_2$ substrate coated with Fe/Mo catalyst nanoparticles \cite{Jie_growth,Jie_growth2}, usually producing nanotubes with diameters of about 2 nm. Individual nanotubes are contacted by metallic leads, thereby forming quantum dots, and controlled by three gates: a back gate that changes the number of electrons in the nanotube and two side gates (SG$_1$ and SG$_2$), located closer to either the source or the drain electrodes. Applying the side gate voltage $V_{\rm SG}$ modifies the relative strength of tunneling from the nanotube to the source and the drain. It turns out that it is sufficient to bias only one of the side gates, as done in this paper.

\begin{figure}[b]
\centering
\includegraphics[width=2.2in]{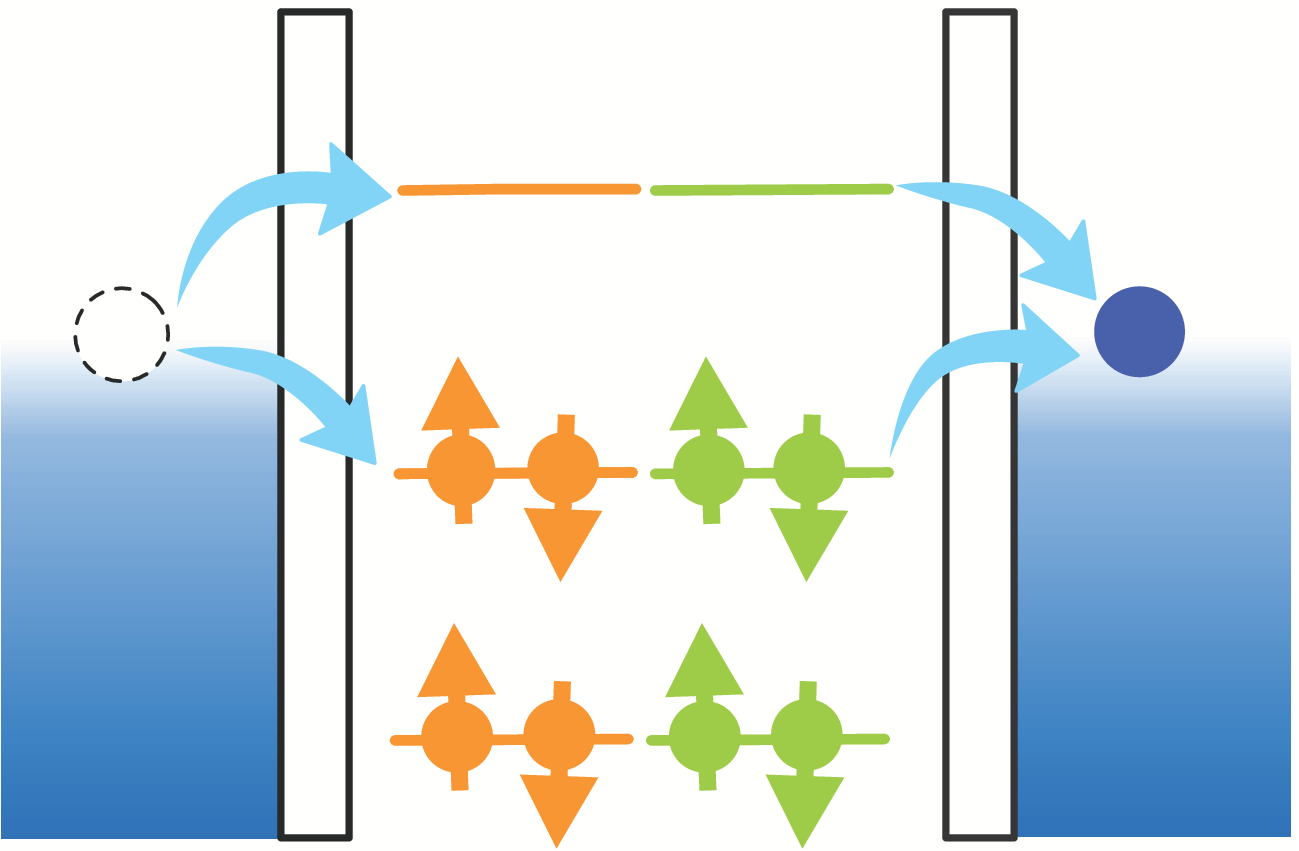} 
\caption{Schematic: an energy diagram of a Coulomb blockade valley, in which an integer number of 4-electron shells is filled (like valleys Y or Z in Figure 1 of the main text). Cotunneling is the dominant process that contributes to the electron transport through the quantum dot: an electron from the source can virtually occupy a high energy orbital in the dot and then tunnel out; alternatively an electron can tunnel out of a filled orbital, followed by an electron from the lead filling the empty state.}
 \label{fig:bdegen}
\end{figure}

Figure 1B of the main text shows a typical plot of the nanotube conductance {\it vs.} gate voltage (the ``Coulomb blockade pattern''). A group of 4 peaks of similar height is visible on the left, corresponding to filling a 4-electron ``shell'' \cite{4peaks, Buitelaar:2002ft}. The peaks in the group are separated from neighboring groups by wider Coulomb blockade valleys Y and Z, in which an integer number of shells is completely filled. In this regime, the electron transport through the nanotube is determined by the co-tunneling processes (Figure S1), which are almost energy-independent on the energy scales smaller than the charging energy or the level spacing ({\it i.e.} meV's). In this case, the nanotube essentially behaves as a lumped tunnel junction. As discussed below and in the main text, this allows us to {\it in situ} characterize the suppression of tunneling conductance due to the resistive leads (Figures 1C-D of the main text).

\section{Dissipative environment}

The strength of dissipation in the resistive leads is characterized by $r=e^2R/h$, where $R$ is the total  resistance of the two leads. The leads are made from a Cr/Au (10nm/1nm ) film with a resistivity of 75 $\Omega/\square$, and their total room temperature resistance is estimated at $\sim$ 6.5 k$\Omega$ for the sample shown in Figure 1 of the main text, and $\sim$ 17 k$\Omega$ for the sample shown in Figures 2-3. These numbers yield $r \sim 0.25$ for the first sample, and $r \sim 0.65$ for the second sample. However the film resistivity increases by at least $10\%$ at low temperature, making these estimates consistent with $r=0.3$ we extract in Figure 1D for the first sample, and $r=0.75$ extracted for the second sample (see the next paragraph).

According to the theory of tunneling with dissipation (also referred to as ``environmental Coulomb blockade'' 
\cite{delsing_1989,cleland_1990,ingoldbook,Flensberg_review,kauppinen_1996,joyez_how_1998,Averin_Lukens}), in order to determine the dissipation strength, one has to consider the impedance of the whole circuit at frequencies corresponding to temperature or bias ({\it i.e.} $\hbar \omega \sim k_BT$ or $eV$). In our case, this frequency range extends from $\sim 1$ GHz up to $\sim 100$ GHz. The lithographically made resistive leads connect the nanotubes to much larger pads (hundreds of microns on a side), whose capacitance short-circuits the high frequency fluctuations. Therefore, only the on-chip resistance of the leads contributes to dissipation. We also estimate that the distributed capacitance of the resistive leads can be neglected in this frequency range, so that the leads behave as simple frequency - independent resistors. 

We choose to demonstrate two different aspects of the observed behavior in two samples. The sample shown in Figure 1 has the smaller dissipation strength of $r=0.3$, resulting in the striking cusp $G \propto \vert V \vert ^{2r}$ in Figure 1C. In Figures 2-4, we extract the peak width proportional to $T^{r/(r+1)}$. The relatively large dissipation strength of this sample is required to distinguish this non-trivial exponent a from the dependence $T^r$ predicted by the lower-order theoretical considerations.

Tunneling current $I(V,T)$ through a single tunneling barrier in the case of Ohmic dissipation was obtained in Ref. \cite{Averin_Lukens}:
\begin{eqnarray}
I(V,T) \propto V T^{2r} {\bigg\vert\frac{\Gamma[r+1+i \frac{eV}{2\pi k_BT}]}{\Gamma[1+i \frac{eV}{2 \pi k_BT}]}\bigg\vert}^2
\label{eq:zheng}
\end{eqnarray}
We numerically differentiate this expression with respect to $V$ to fit the data in Figure 1d of the main text.

\section{Estimates of the exponents}

{\it Asymmetric peak height.} Figure 3A shows the temperature dependence of the peak height for various degrees of asymmetry between the tunneling rate from the level to the source and the drain electrodes, $\Gamma_S$ and $\Gamma_D$. Depending on the asymmetry, the peak conductance exhibits a range of cross-over behaviors. For peaks with low degree of asymmetry (such that $|\Gamma_S- \Gamma_D| \sim k_B T \ll |\Gamma_S+ \Gamma_D|$, top curves), the proper temperature scaling of conductance is not yet fully developed, so care must be taken to avoid extracting an incorrect exponent. 
Hence we study the conductance scaling of the most asymmetric peak of Figure 3A (reproduced here in Figure S2), which demonstrates a fully developed power-law behavior in the accessible temperature range. A linear curve is fitted to the data on a log-log scale in the temperature range of 0.08 K $<T<$ 1.2 K. The extracted slope of 1.47 $\pm$ 0.04 corresponds to the value of $r = 0.74 \pm 0.02$. 

\begin{figure}[h]
 \centering
 \includegraphics[width=3.3in]{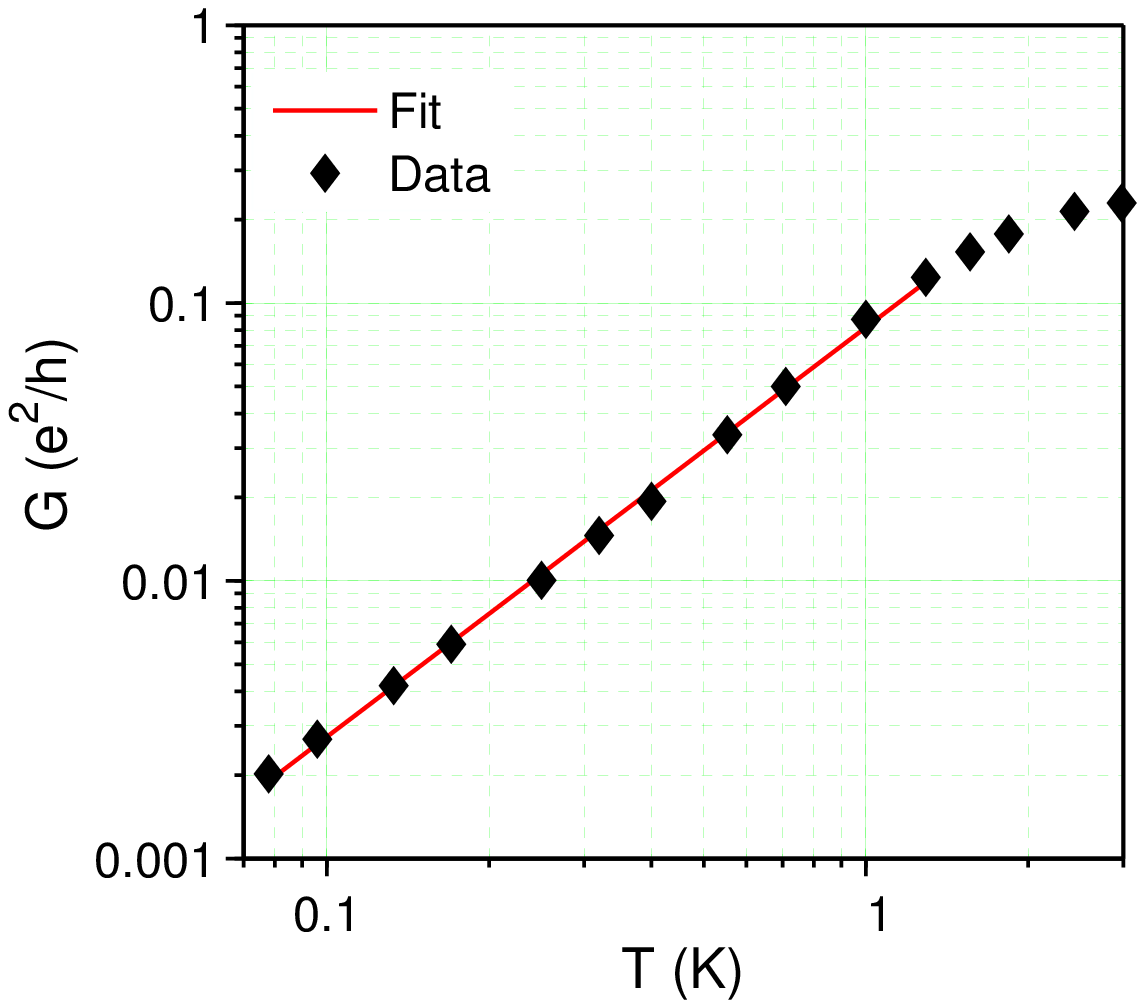} 
 \caption{Conductance of an asymmetrically coupled peak {\emph vs.} temperature (same data as the lowest curve in Figure 3A), and the power-law fit. The uncertainty in the data points is about the symbol size.}
 \label{fig:asymmetricG}
\end{figure}

{\it Symmetric resonance width.} 
The full-width at half-maximum (FWHM) of the symmetric peak (Figure 2A) is plotted in Figure S3. The fit shown in red yields an exponent of 0.43 $\pm$ 0.01, surprisingly close to the expected value of $r/(r+1) \approx 0.43$. (In order to avoid over-estimating the exponent, the $T> 1$ K data points are not included in the fit. This higher temperature range corresponds to the transition to the sequential tunneling regime $k_B T \gtrsim \Gamma$, in which case the peak width becomes proportional to $T$.)   

\begin{figure}[h]
 \centering
 \includegraphics[width=3.1in]{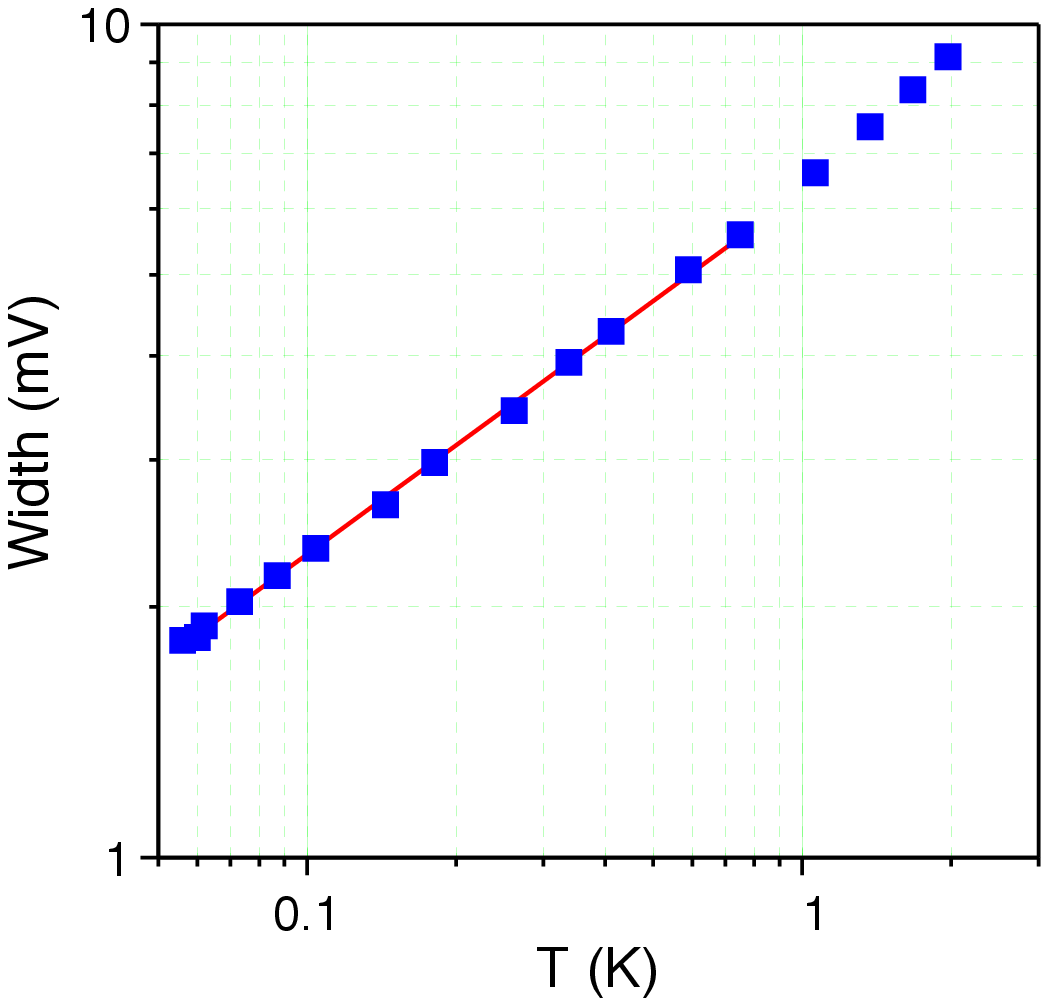} 
\caption{Full-width at half-maximum extracted from the symmetric peak in Figure 2B. In order to exclude the contribution from the sequential tunneling regime $k_B T \gtrsim \Gamma$, only the data points measured at $T < 1$ K are used for the fit shown in red (see text). The vertical axis has to be multiplied by the experimentally determined "gate efficiency factor" of $\approx$ 0.2, which converts $\Delta V_{gate}$ units to the actual energy of the resonant level (see e.g. Ref \cite{QDreview}). }
 \label{fig:symmetricFWHM}
\end{figure}

Before proceeding to the theoretical model, we note that our observations cannot be explained by the conventional Lorentzian expression for the resonance conductance between non-interacting leads \cite{Kastner}. First, the line shape of the resonances is not Lorentzian. Second, even if one tries to describe the symmetric case by a Lorentzian with temperature-dependent tunneling rates $\Gamma_{S,D}(T)$, in the asymmetric case the same expression would also yield a peak with a temperature-independent height and a vanishing width, in marked contrast with our measurements in Fig.~2c.

\section{Model}

\begin{figure}[h]
\centering
\includegraphics[width=0.6 \columnwidth]{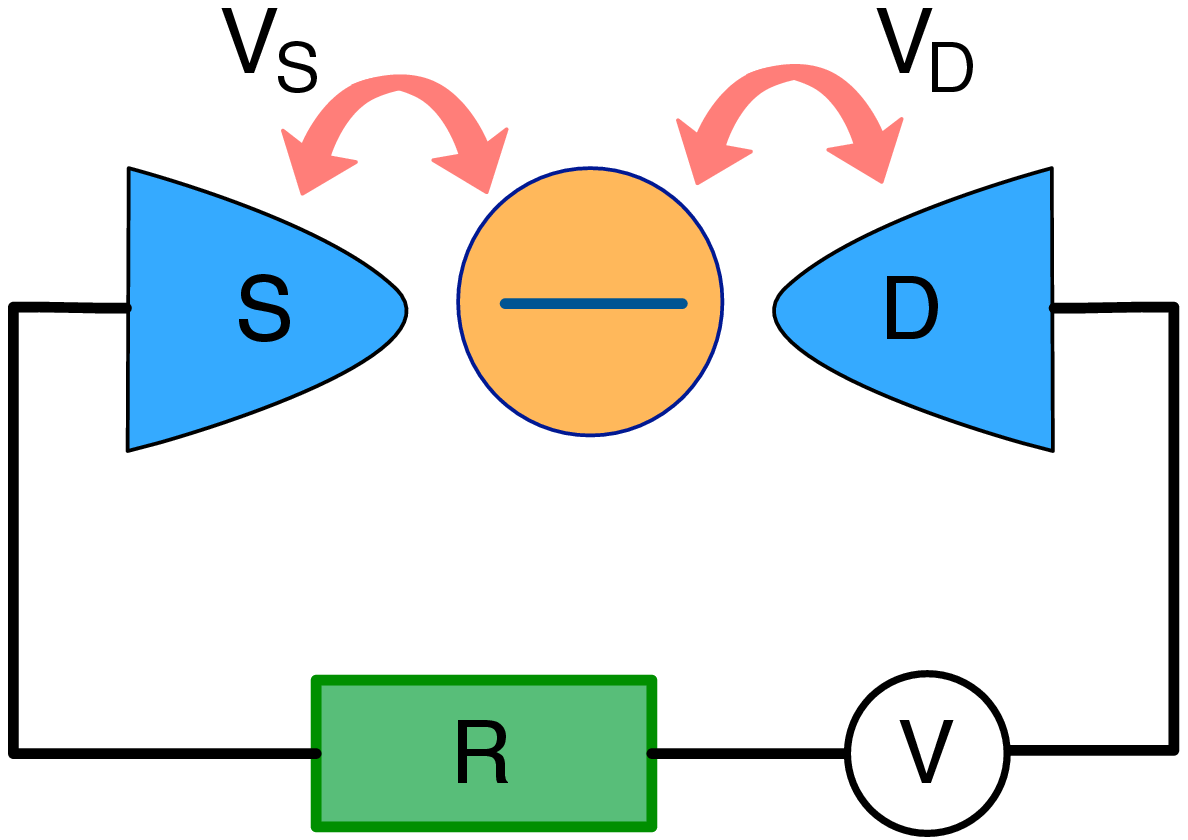}
\caption{Schematic: a spinless quantum dot is coupled to two conducting leads with tunneling amplitudes $V_{S}$ and $V_{D}$. The dot-leads system is attached to a voltage source $V$ via $R$, the sum of the lead resistances. }
\end{figure}

We model our experimental situation by a spinless level coupled to two conducting leads in the presence of an Ohmic dissipative environment (Figure S4). As we show in the following, the electromagnetic modes of the environment mediate interactions between the electrons in the leads, resulting in a Luttinger liquid-like behavior. We start with a system Hamiltonian given by
\begin{equation}
H=H_{\textrm{Dot}}+H_{\textrm{Leads}}+H_{\textrm{T}}+H_{\textrm{Env}}
\label{eq:H}
\end{equation}
and describe the individual terms in this section.
$H_{\textrm{Dot}}=\epsilon_{\textrm{d}}d^{\dagger}d$ is the Hamiltonian of the dot with the energy level $\epsilon_{\textrm{d}}$ and the electron creation operator $d^{\dagger}$. $H_{\textrm{Leads}}=\sum_{\textrm{\textrm{\ensuremath{\alpha}=S,D}}}\sum_{k}\epsilon_{k}c_{k\alpha}^{\dagger}c_{k\alpha}$
represents the electrons in the source (S) and drain (D) leads. 

$H_{\textrm{T}}$ describes the tunneling between the dot and the leads:
\begin{equation}
H_{\textrm{T}}=V_{S}\sum_{k}(c_{kS}^{\dagger}e^{-i\varphi_{S}}d+  )+V_{D}\sum_{k}(c_{kD}^{\dagger}e^{i\varphi_{D}}d+ {\rm h.c.} ),
\end{equation}
where the operators $\varphi_{S/D}$ represent the phase fluctuations of the tunneling amplitude between the dot and the S/D lead. These phase operators are canonically conjugate to the operators $Q_{S/D}$ corresponding to charge fluctuations on the S/D junctions. This is a standard way to treat macroscopic quantum tunneling in the presence of a dissipative environment \cite{ingoldbook}, which is valid for electrons propagating much slower
than the electromagnetic field \cite{Nazarovbook}. 

It is useful to transform to variables related to the total charge on the dot. To that end, we introduce \cite{ingoldbook} two new phase operators, $\varphi$ and $\psi$, related to the phases $\varphi_{S/D}$ 
by 
\begin{eqnarray}
\varphi_{S} & = & \kappa_{S}\varphi+\psi \qquad\nonumber \\
\varphi_{D} & = & \kappa_{D}\varphi-\psi \;,\label{eq:}
\end{eqnarray}
where $\kappa_{S/D}=C_{D/S}/(C_{S}+C_{D})$ in terms of the capcitances of the two dots, $C_{S/D}$. $\psi$ is the variable conjugate to the fluctuations of charge on the dot $Q_{c}=Q_{S}-Q_{D}$ and so couples to voltage fluctuations on the gate which controls the energy of the dot's level. Likewise, $\varphi$ is the variable conjugate
to $Q=(C_{S}Q_{D}+C_{D}Q_{S})/(C_{D}+C_{S})$. Assuming $C_{S}=C_{D}$, we have
$\varphi_{S}=\varphi/2+\psi$ and $\varphi_{D}=\varphi/2-\psi$. 

The gate voltage fluctuations can be disregarded in our experiment because the capacitance of the gate is negligible, $C_{g}\ll C_{S/D}$. (The opposite limit of a noisy gate coupled to a resonant level was considered in Ref.\,\onlinecite{Averin_94}.) In fact, for our purposes, the coupling of the fluctuations of the total charge on the dot to the environment can be neglected. Thus, only the {\it relative} phase difference between the two leads remains \cite{ingoldbook,florens07}, and the tunneling Hamiltonian becomes
\begin{equation}
H_{\textrm{T}}=V_{S}\sum_{k}(c_{kS}^{\dagger}e^{-i\frac{\varphi}{2}}d+ {\rm h.c.})+V_{D}\sum_{k}(c_{kD}^{\dagger}e^{i\frac{\varphi}{2}}d+ {\rm h.c.} ). 
\end{equation}

The last part of Eq. (\ref{eq:H}) is the Hamiltonian of the environment, $H_{\textrm{Env}}$ \cite{caldeira81,leggett87,ingoldbook}. The environmental modes are represented by harmonic oscillators described by inductances and capcitances such that their frequencies are given by 
$\omega_{n}=1/\sqrt{L_{n}C_{n}}$. These oscillators are then bilinearly coupled to the phase operator $\varphi$ through the oscillator phase:
\begin{equation}
H_{\textrm{Env}}=\frac{Q^{2}}{2C}+\sum_{n=1}^{N}\left[\frac{q_{n}^{2}}{2C_{n}}+\left(\frac{\hbar}{e}\right)^{2}\frac{1}{2L_{n}}\left(\varphi-\varphi_{n}\right)^{2}\right]. 
\end{equation}

\section{Mapping to Luttinger Liquids}

In this section, we demonstrate the mapping of our model to that of a resonant level contacted by two Luttinger liquids. In carrying out this mapping, we follow closely previous work on tunneling through a single barrier with an environment \cite{safi04,LeHur&Li_05} and the Kondo effect in the presence of resistive leads \cite{florens07}. 

The two metallic leads in our case can be reduced to two semi-infinite one-dimensional free fermionic baths, which are non-chiral \cite{GiamarchiBook}. By unfolding them, one can obtain two chiral fields \cite{GiamarchiBook}, which both couple to the dot at $x=0$. We bosonize the fermionic fields in the standard way \cite{GiamarchiBook,senechal} $c_{S/D}(x)=\frac{1}{\sqrt{2\pi a}}F_{S/D}\exp[i\phi_{S/D}(x)]$. Here, $F_{S/D}$ is the Klein factor, $\phi_{S/D}$ is the bosonic field, and $a$ is the short time cutoff. Defining the flavor field $\phi_{f}$ and charge field $\phi_{c}$ by
\begin{equation}
\phi_{f} \equiv \frac{\phi_{S}-\phi_{D}}{\sqrt{2}}, \qquad
\phi_{c} \equiv \frac{\phi_{S}+\phi_{D}}{\sqrt{2}},
\end{equation}
we rewrite the Hamiltonian of the leads as
\begin{equation}
H_{\textrm{Leads}}=\frac{v_{F}}{4\pi}\int_{-\infty}^{\infty}dx\left[\left(\partial_{x}\phi_{c}\right)^{2}+[\left(\partial_{x}\phi_{f}\right)^{2}\right]. 
\end{equation}
The tunneling Hamiltonian then becomes
\begin{widetext}
\begin{eqnarray}
H_{\textrm{T}} & = & V_{S}\left(\frac{1}{\sqrt{2\pi a}}F_{S}\,\text{exp}\left[-i\frac{\phi_{c}(0)+\phi_{f}(0)}{\sqrt{2}}\right]e^{-i\frac{\varphi}{2}}d+ {\rm h.c.} \right)\nonumber \\
 & + & V_{D}\left(\frac{1}{\sqrt{2\pi a}}F_{D}\,\text{exp}\left[-i\frac{\phi_{c}(0)-\phi_{f}(0)}{\sqrt{2}}\right]e^{i\frac{\varphi}{2}}d+ {\rm h.c.} \right). 
\end{eqnarray}
Note a key feature of $H_{\textrm{T}}$: the fields $\varphi$ and $\phi_{f}(0)$ enter in the same way. Thus we wish to combine these two fields, a process which will lead to effectively interacting leads as in a Luttinger liquid. 

To carry out such a combination, since the tunneling only acts at $x=0$, it is convenient to perform
a partial trace in the partition function and integrate out fluctuations in $\phi_{c/f}(x)$ for all $x$ away from $x=0$ \cite{KF}. For an Ohmic environment, one can also integrate out the harmonic modes \cite{leggett87,safi04,florens07}. Then, the effective action for the leads and the environment becomes
\begin{equation}
S_{\textrm{Leads+Env}}^{\textrm{eff}}=\frac{1}{\beta}\sum_{n}|\omega_{n}|\left(|\phi_{c}(\omega_{n})|^{2}+|\phi_{f}(\omega_{n})|^{2}+\frac{R_{Q}}{2R}|\varphi(\omega_{n})|^{2}\right),
\end{equation}
where $R_{Q}=h/e^{2}$, $R$ is the total resistance of the leads,
and $\omega_{n}=2\pi n/\beta$ is the Matsubara frequency. 
In this discrete representation, it is straightforward to combine 
the phase operator $\varphi$ and the flavor field $\phi_{f}$; in order to maintain canonical commutation relations while doing so, we use the transformation
\begin{equation}
\phi_{f}' \equiv \sqrt{g_{f}}\left(\phi_{f}+\frac{1}{\sqrt{2}}\varphi\right),\qquad
\varphi' \equiv \sqrt{g_{f}}\left(\sqrt{\frac{R}{R_{Q}}}\phi_{f}-\sqrt{\frac{R_{Q}}{R}}\frac{1}{\sqrt{2}}\varphi\right),
\end{equation}
where $g_{f}\equiv 1/(1+R/R_{Q})<1$. Now, the effective action
for the leads and environment (excluding tunneling) becomes
\begin{equation}
S_{\textrm{Leads+Env}}^{\textrm{eff}}=\frac{1}{\beta}\sum_{n}|\omega_{n}|\left(|\phi_{c}(\omega_{n})|^{2}+|\phi_{f}'(\omega_{n})|^{2}+|\varphi'(\omega_{n})|^{2}\right) \;,
\label{eq:Seff}
\end{equation}
while the Lagrangian for the tunneling part reads 
\begin{equation}
L_{\textrm{T}}=-V_{S}\left(\frac{F_{S}}{\sqrt{2\pi a}}e^{-i\frac{1}{\sqrt{2g_c}}\phi_{c}(\tau)}e^{-i\frac{1}{\sqrt{2g_{f}}}\phi_{f}'(\tau)}d+{\rm c.c.}\right)-V_{D}\left(\frac{F_{D}}{\sqrt{2\pi a}}e^{-i\frac{1}{\sqrt{2g_c}}\phi_{c}(\tau)}e^{i\frac{1}{\sqrt{2g_{f}}}\phi_{f}'(\tau)}d+ {\rm c.c.}\right) \;. 
\label{eq:LT}
\end{equation}
\end{widetext}
(Here, we have formally introduced the parameter $g_{c}=1$ to describe the noninteracting field $\phi_{c}$.) Indeed, we see that the phase $\varphi$ has been absorbed into the new flavor field $\phi_{f}'$ at the expense of a modified interaction parameter $g_{f}$, while the new phase fluctuation $\varphi'$ decouples from the system. In what follows, we will drop the prime from the operator $\phi_{f}'$. 

It turns out that one obtains a very similar effective action by starting from a model with resonant tunneling between Luttinger liquids \cite{KF,note3}. The two models are equivalent if the interaction parameters $g_{c}$ and $g_{f}$ in our model are made equal to the single interaction parameter $g$ of Ref.~\onlinecite{KF}. Similar mappings were obtained for a spinful model in the absence of charge fluctuations (Kondo regime) in Ref.~\onlinecite{florens07} and for a dissipative dot coupled to a single chiral Luttinger liquid in Ref.~\onlinecite{LeHur&Li_05}.

As discussed in the main text, the $r=0$ case ($g_f=1$) reduces to a resonant level without dissipation, while the $r=1$ case ($g_f=1/2$) can be mapped onto the two-channel Kondo model. To understand the relation between our model and the two-channel Kondo model, we apply a unitary transformation \cite{emery92,komnik03}, $U=\exp[i(d^{\dagger}d-1/2) \phi_c(0) / \sqrt{2}]$, to eliminate the $\phi_c$ field in the tunneling Lagrangian, Eq. (\ref{eq:LT}). At the same time, an extra electrostatic Coulomb interaction between the leads and the dot is generated. Introducing in addition a bare electrostatic Coulomb interaction ($U_b$), we have 
\begin{equation}
 H_{C}=\frac{\sqrt{2}}{\pi} (U_b-1) (d^{\dagger}d-1/2) \partial_x \phi_c(x=0) \;.
\end{equation}
For the special value $g_f=1/2$, we can refermionize the problem by defining $\psi_{c,f}=e^{i\phi_{c,f}}/\sqrt{2\pi a}$. 
Then, at $V_S=V_{D}$ (and $r=1$), our model is mapped onto the two-channel Kondo model \cite{emery92,komnik03,goldstein10,Affleck}, which shows non-Fermi-liquid behavior \cite{GogolinBook}.
At the Toulouse point ($U_b=1$ so that $H_{C}=0$), the model reduces to a Majorana resonant level 
model \cite{emery92,GogolinBook}.
Finally, for $r$ close to $1$ (\emph{ i.e.} in our case), one can similarly
map our system onto a two-channel Kondo model with (interacting)
Luttinger liquid leads \cite{goldstein10,Affleck}. The effective interaction parameter $g_\sigma$
in this case is determined by the residual $1-r$. 

 \vspace{\baselineskip} 
\section{Quantum Phase Transition and Scaling Relations }

Having established the relation between our problem and Luttinger liquid physics, we can draw on the very extensive theoretical work concerning resonant tunneling in a Luttinger liquid \cite{KF,Eggert&Affleck1992,Sassetti_Napoli_Weiss_95,nazarov03,polyakov03,komnik03,Meden2005} to reach conclusions about the scaling and phase transitions implied by our model.
To understand how the interacting environment affects the low temperature
physics, it is convenient to rewrite the model, following Refs.~\onlinecite{KF} and \onlinecite{Goldstein10a}, in the ``Coulomb-gas'' representation, which can be accomplished by expanding
the partition function in powers of $V_{S}$ and $V_{D}$. After integrating
out $\phi_{f}(\tau)$ and $\phi_{c}(\tau)$ in each term, one obtains
a classical one-dimensional statistical mechanics problem with the
partition function 
\begin{widetext}
\begin{eqnarray}
Z & = & \sum_{\sigma=\pm}\sum_{n}\sum_{\{q_{i}=\pm\}}V_{S}^{\sum_{i}(1+q_{i}p_{i})/2}V_{D}^{\sum_{i}(1-q_{i}p_{i})/2}
\nonumber \\ & & \times
 \intop_{0}^{\beta}d\tau_{2n}\intop_{0}^{\tau_{2n}}d\tau_{2n-1}. . . . . . \intop_{0}^{\tau_{2}}d\tau_{1}\exp\Big\{\sum_{i<j}V_{ij}\Big\}\:\exp\left\{\epsilon_{d} \Big[\beta\frac{1-\sigma}{2}+\sigma\sum_{1\leq i\leq2n}p_{i}\tau_{i}\Big]\right\},\nonumber \\
V_{ij} & = & \frac{1}{2g_{f}}\left[q_{i}q_{j}+K_{1}p_{i}p_{j}+K_{2}(p_{i}q_{j}+p_{j}q_{i})\right]\ln\left(\frac{\tau_{i}-\tau_{j}}{\tau_{c}}\right) \;. 
\end{eqnarray}
\end{widetext}
We consider the on-resonance case, $\epsilon_{d}=0$, so that the last term in the partition function is equal to $1$. There are two types of charges in this 1D problem \cite{KF}:
$q_{i}$ charge and $p_{i}$ charge, both of which can be $\pm1$. 
The total system is charge neutral, $\sum_{i}q_{i}=\sum_{i}p_{i}=0$. 
Physically, the $q_{i}$ charge corresponds to a tunneling event
between the dot and the leads ($+1$ : to the right (D); $-1$: to the
left (S) ). The $p_{i}$ charge corresponds to hopping onto ($+1$) or off
($-1$) the dot. 

Following the method developed for resonant tunneling between Luttinger liquids in Ref.~\onlinecite{KF}, one obtains the renormalization group (RG) equations and the phase diagram. The sign of the $p_{i}$ charge must alternate in time, while the $q_{i}$ charge can have any ordering satisfying the charge neutrality constraint. Therefore, the interaction between the $q_{i}$ charges does not vary in the RG flow, while the interaction strength between $p_{i}$ charges, $K_{1}$, does renormalize. The bare $K_{1}$ (initial value in RG flow) is $K_{1}^{\text{bare}}=g_{f}/g_{c}$; in fact, by taking into account the coupling of the fluctuations of the total dot charge to the environment---an effect neglected here---one can show that $K_{1}^{\text{bare}}= 1$ \cite{Dong2012}. The bare value of $K_{2}$ is zero, but it will be generated in the RG flow. These are the same conditions as in resonant tunneling between two Luttinger liquids \cite{KF}. Hence, our model is mapped onto the model of resonant tunneling between Luttinger liquids, with the RG equations for the interaction strengths and tunneling couplings as in Ref.~\onlinecite{KF}.

\begin{figure}[t]
\includegraphics[scale=0.6]{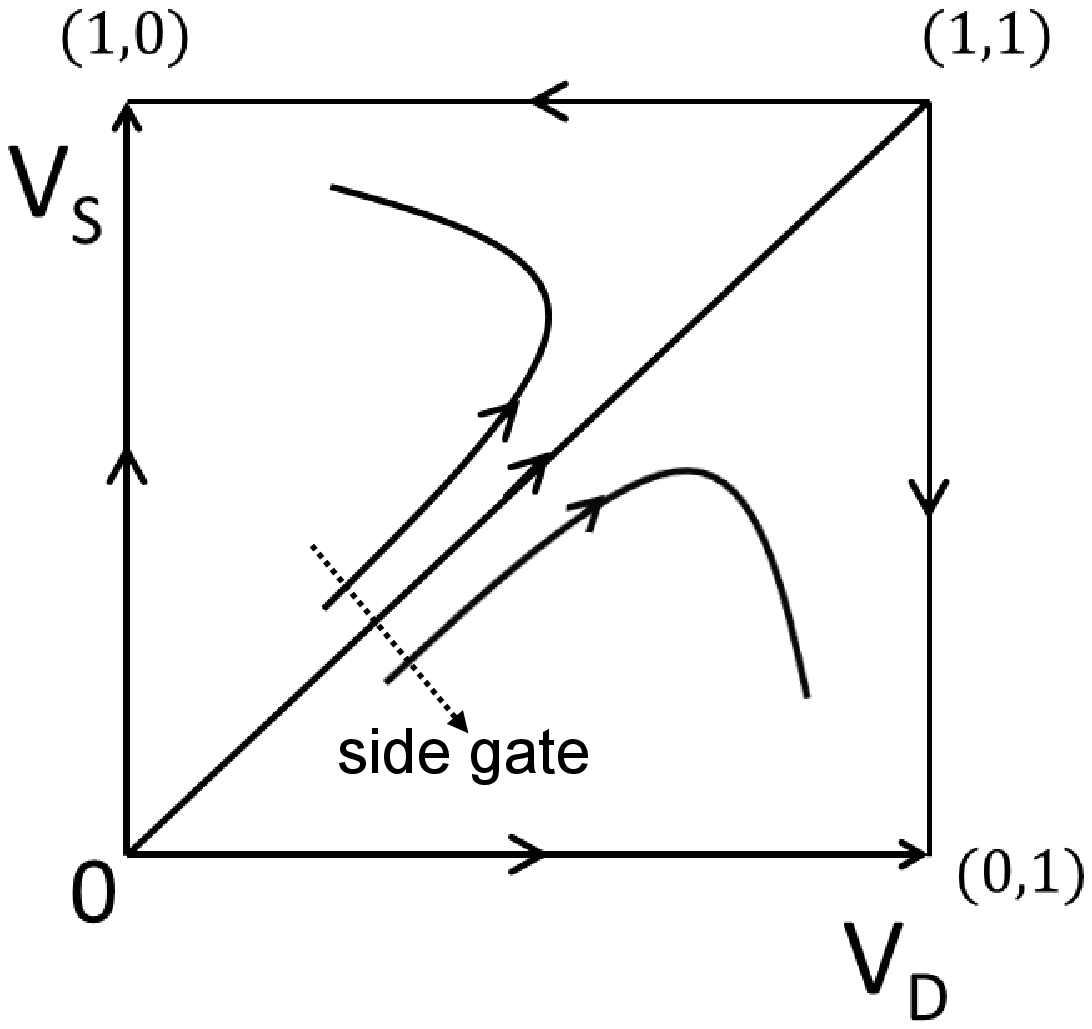}
\caption{Schematic representation of the renormalization group flow for the two tunneling amplitudes, $V_{S}$ and $V_{D}$ \cite{KF}. The diagonal corresponds to the symmetric coupling case, and flows into the strongly coupled quantum critical point at (1,1) which corresponds to a single homogeneous Luttinger liquid. Point (1,0) describes the strong coupling point for $V_{S}$, while $V_{D}=0$; similarly, point (0,1) describes the strong coupling point for $V_{D}$, while $V_{S}=0$. }
\end{figure}

The resulting schematic RG flow diagram in the $V_S$-$V_D$ plane, valid for  $r<1$ and on-resonance, is shown in Figure S5.
A remarkable feature of this RG flow diagram is that a second order quantum phase transition can be realized by tuning the tunneling matrix element across the symmetric coupling line, $V_{S} = V_{D}$. Indeed, this critical value separates the flows terminating in the stable fixed points denoted as $(0,1)$ and $(1,0)$ in Figure S5. These two phases correspond to the dot merging with the D lead, while the S lead decouples, or {\it vice versa}. The unstable fixed point denoted as $(1,1)$ is reached starting exactly at $V_{S} = V_{D}$; it corresponds to a \textit{homogeneous} Luttinger liquid which therefore has conductance $G=e^{2}/h$ \cite{KF,Eggert&Affleck1992}. 

The proximity to the strongly coupled point $(1,1)$ determines the quantum critical behavior and the critical exponents observed in the experiment. By analyzing the universal scaling function \cite{KF}, the width of the resonant peak is found to scale to zero as $\Gamma\propto T^{1-g_{f}}=T^{\, r/(1+r)}$. For the case of asymmetric tunneling $V_{S}\neq V_{D}$, either $V_{S}$ or $V_{D}$ flows to zero so that the height of the resonant peak scales as $G\propto T^{2\left(1/g_{f}-1\right)=}=T^{2r}$ \cite{KF}, while the width of the resonance saturates \cite{nazarov03,polyakov03,komnik03}. 

These power-law dependencies are typical for the critical behavior near a second-order phase transition. We would like to stress that this behavior qualitatively differs from the Kosterlitz-Thouless (KT) type of transition, commonly encountered in quantum impurity models \cite{Vojta}. For example, several recent theoretical works predict KT transitions in quantum dots coupled to a single interactive lead \cite{FurusakiMatveev02,QPT,Borda_06,ChungQPT,ChengIngersent09}. There, the QPT occurs when the dissipation exceeds a certain critical value. In our case, the crucial ingredient that enables the QPT is the symmetric coupling to the two leads, which allows for their competition, while the dissipation strength can be relatively weak (but greater than zero).


\begin{thebibliography}{99}
\bibitem{note1_1} We stress that our experiment does not probe the Luttinger liquid in the nanotube: at this temperature range, the length of an ideally clean nanotube would have to be $\sim 100 \mu$m to suppress the size quantization.
\bibitem{note0_1} The observed behavior cannot be explained by the conventional Lorentzian expression for the resonance conductance between non-interacting leads \cite{Kastner_1}. First, the line shape of the resonances is not Lorentzian. Second, even if one tries to describe the symmetric case by a Lorentzian with temperature-dependent tunneling rates $\Gamma_{S,D}(T)$, in the asymmetric case the same expression would also yield a peak with a temperature-independent height and a vanishing width, in marked contrast with our measurements in Fig.~2c.
\end{thebibliography}
\end{document}